\documentclass[12pt]{article}
\usepackage{graphicx}

\def\pbnr{}
\def\speaker{Alexey A Petrov}

\def\title{Long-distance effects in charm mixing}
\def\affiliation{Department of Physics and Astronomy\\
Wayne State University \\
Detroit, MI 48201 USA}
\def\support{This work was supported in part by the U.S. Department of Energy under contract DE-FG02-12ER41825.}

\newcommand\pubnumber{\pbnr}
\newcommand\pubdate{\today}
\def\Title#1{\begin{center} {\Large #1 } \end{center}}
\def\Author#1{\begin{center}{ \sc #1} \end{center}}

\newcommand{\OnBehalf}[1]{\sbox0{#1}\ifdim\wd0=0pt
        {}
	\else
	{\\on behalf of #1}
	\fi}
\newcommand{\SupportedBy}[1]{\sbox0{#1}\ifdim\wd0=0pt
        {}
	\else
	{\footnote{#1}}
	\fi}
\def\Address#1{\begin{center}{ \it #1} \end{center}}

\newcommand\pubblock{\includegraphics[width=5cm]{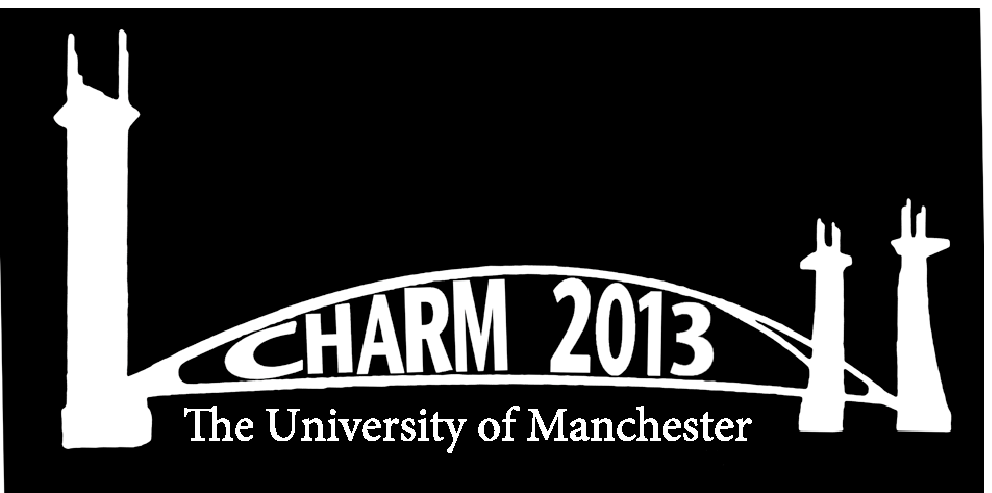}\hfill{\begin{tabular}{l} \pubnumber\\
         \pubdate  \end{tabular}}}
\newenvironment{Abstract}{\begin{quotation}  }{\end{quotation}}
\newenvironment{Presented}{\begin{quotation} \begin{center} 
             PRESENTED AT\end{center}\bigskip 
      \begin{center}\begin{large}}{\end{large}\end{center} \end{quotation}}
\def\Acknowledgements{\bigskip  \bigskip \begin{center} \begin{large}
             \bf ACKNOWLEDGEMENTS \end{large}\end{center}}
\def\venue{The 6$^{th}$ International Workshop on Charm Physics\\
(CHARM 2013)\\
Manchester, UK,  31 August -- 4 September, 2013}

\def\beq{\begin{equation}}
\def\eeq#1{\label{#1}\end{equation}}
\def\eeqn{\end{equation}}

\def\beqa{\begin{eqnarray}}
\def\eeqa#1{\label{#1}\end{eqnarray}}
\def\eeqan{\end{eqnarray}}

\let\bar=\overbar

\def\D{{\cal D}}

\def\Dslash{\not{\hbox{\kern-4pt $D$}}}
\def\dslash{\not{\hbox{\kern-2pt $\del$}}}

\def\msb{{\bar{\ssstyle M \kern -1pt S}}}

\def\barD{\overline D{}^0}

\def\DDbar{D^0-\overline D{}^0}

\def\D0bar{\overline D{}^0}
\def\K0bar{\overline K{}^0}

\def\3bar{\overline{3}}
\def\sixbar{\overline{6}}

\def\15bar{\overline{15}}
\def\24bar{\overline{24}}
\def\42bar{\overline{42}}
\def\60bar{\overline{60}}
\def\cO{{\it O}}
\def\cal{{\it}}

\def\beq{\begin{equation}}
\def\eeq{\end{equation}}
\def\beqa{\begin{eqnarray}}
\def\eeqa{\end{eqnarray}}
\def\bea{\begin{eqnarray}}
\def\eea{\end{eqnarray}}

\begin{document}
\begin{titlepage}
\pubblock

\vfill
\Title{\title}
\vfill
\Author{\speaker\SupportedBy{\support}}
\Address{\affiliation}
\vfill
\begin{Abstract}
I review challenges in the understanding of long distance effects in theoretical calculations
of mixing rates of charmed mesons in the standard model.
\end{Abstract}
\vfill
\begin{Presented}
\venue
\end{Presented}
\vfill
\end{titlepage}
\def\thefootnote{\fnsymbol{footnote}}
\setcounter{footnote}{0}
%

\section{Introduction}

Flavor physics, especially physics of charmed mesons, offers incredibly rich opportunities 
not only to study soft Quantum Chromodynamics (QCD), but also search for glimpses of new 
physics (NP)~\cite{Artuso:2008vf,Golowich:2007ka}. That search is only possible if the 
standard model (SM) predictions for experimental observables are known well, which means that
uncertainties of theoretical predictions are understood and under control. The experimental 
observables, such as meson mixing parameters, rates and asymmetries in rare decays and/or
CP-violating asymmetries, are designed to provide likely places where NP can be 
observed~\cite{Artuso:2008vf}. Among those, a steady improvement of precision of 
experimental observation of $\DDbar$ mixing rate, offers a great hope that possible 
NP contributions in the up-quark sector would be soon constrained or observed~\cite{Exp}.
Unfortunately, quantitative theoretical understanding of $\DDbar$ mixing rate remains one of 
the most difficult problems in flavor physics. 

The $\Delta C = 2$ interactions, generated either at one loop level in the SM or possibly
by NP particles, mix a $D^0$ state into a $\barD$ state, which 
results in physical (measurable) mass and lifetime differences between new
mass eigenstates~\cite{BigiSandaBook},
\begin{eqnarray} \label{definition}
x_D \equiv \frac{m_2-m_1}{\Gamma}, ~~
y_D \equiv \frac{\Gamma_2 - \Gamma_1}{2 \Gamma},
\end{eqnarray}
where $m_{1,2}$ and $\Gamma_{1,2}$ are the masses and widths of
$D_{1,2}$  and the mean width and mass are $\Gamma=(\Gamma_1+\Gamma_2)/2$ 
and $m=(m_1+m_2)/2$. The mass eigenstates themselves are usually defined as
\begin{equation} \label{definition1}
| D_{^1_2} \rangle =
p | D^0 \rangle \pm q | \bar D^0 \rangle,
\end{equation}
where the complex parameters $p$ and $q$ are obtained from diagonalizing 
the $D^0-\barD$ mass matrix. The mass and lifetime differences introduced above 
can be calculated as absorptive and 
dispersive parts of a certain correlation function,
\begin{eqnarray}\label{XandY}
x_D &=& \frac{1}{M_D\Gamma_D} \mbox{Re}
\Biggl[
2 \langle \D0bar | {\cal H}_w^{|\Delta C| = 2} | D^0\rangle 
\nonumber \\
&+& 
 \langle \D0bar | 
 i \int d^4x ~\mbox{T}\left\{ {\cal H}_w^{|\Delta C|=1} (x)  {\cal H}_w^{|\Delta C|=1} (0)\right\}
 | D^0\rangle 
 \Biggr],
 \\
y_D &=& \frac{1}{M_D\Gamma_D} \mbox{Im} \left[
 \langle \D0bar | 
 i \int d^4x ~\mbox{T}\left\{ {\cal H}_w^{|\Delta C|=1} (x)  {\cal H}_w^{|\Delta C|=1} (0)\right\}
 | D^0\rangle \right].
\end{eqnarray}
It is understood that only quarks whose masses are lighter than $m_D$ can go on mass shell in 
Eq.~(\ref{XandY}) and provide nonzero value for the lifetime difference $y_D$.

Charm system is quite unique because $x_D$ is {\it not} dominated by the contribution of the 
$\Delta C = 2$ operator that is local at the charm scale. This is very different from the case of 
$B$-mixing, where $x$ is completely dominated by the top quark contribution. Since 
Glashow-Iliopoulos-Maiani (GIM) guarantees that the mixing amplitude is proportional to the power
of intrinsic quark mass running in the box diagram, suppressions due to a combination of 
Cabibbo-Kobayash-Maskawa (CKM) greatly diminish the contribution due to $b$-quark, the only 
heavy quark intermediate state possible in $\DDbar$ mixing. Thus, it is absolutely important to 
calculate the contribution due to the correlation functions in Eq.~(\ref{XandY}) with light
intermediate $s$ and $d$ quarks.

The hardest problem in charm mixing is to properly evaluate the integrals in the above equations.
This can be done in several ways, depending on whether one considers the decaying particle as 
heavy or light compared to the QCD's scale $\Lambda_{QCD}$. Since $m_c \simeq 1.3$~GeV,
both approaches are possible for $D$-decays and mixing calculations.

If the decaying particle is heavy, it is possible to show~\cite{Shifman:1986sm} that 
the integrals in Eq.~(\ref{XandY}) are dominated by short distances, so a 
short-distance operator product expansion (OPE) can be used to evaluate 
the products of $|\Delta C|=1$ Hamiltonians. Similar approaches worked very well
for the calculations of lifetime differences of $B_s$ mesons~\cite{Beneke:1996gn}.
If the decaying particle is considered light, no short-distance expansion of operator products 
is possible, as the integrals are dominated by the long distances. However, 
only a few open channels are available for such light particles, so the 
calculations can be done by explicitly summing over the contributions from each 
of the channels. This approach worked well for kaon physics\footnote{It is important to
remember that this statement only refers to the bilocal part of the expressions for $x$ and $y$. 
The mass difference in kaons is dominated by the contribution from heavy $t$ and $c$ quarks, i.e.
by the ${\cal H}_w^{|\Delta C| = 2}$.}. The number of available decay channels is quite large, but 
some predictions can nevertheless be made.

\section{Expectations for $x_D$ and $y_D$}

Before proceeding to the calculation of $\DDbar$ mixing amplitude, let us understand the 
underlying flavor symmetry structure. GIM mechanism implies that meson mixing amplitudes
must be proportional to mass factors of quarks propagating in the loops providing $\Delta C=2$ 
interactions. Neglecting for a moment the third generation, only $s$ and $d$ quarks give
contribution to $x_D$ and $y_D$ in the standard model. This means that GIM mechanism implies 
that $\DDbar$ mixing is an $SU(3)_F$-breaking effect, and predicting the
Standard Model values of $x_D$ and $y_D$ depends crucially on estimating the 
size of $SU(3)_F$ breaking.  The question is at what order in $SU(3)_F$-breaking
parameter $m_s$ does the effect become non-zero?

To answer that, let us look at the group-theoretical structure of mixing matrix elements
$\langle\D0bar|\, {\cal H}_w {\cal H}_w\, |D^0\rangle\,$ that define $x_D$ and $y_D$.
Here ${\cal H}_w$ denote the $\Delta C=-1$ part of the weak Hamiltonian.  
Let $D$ be the field operator that creates a $D^0$ meson and annihilates a
$\D0bar$.  Then the matrix element, whose $SU(3)$ flavor group theory properties
we will study, may be written as
\beq\label{melm}
      \langle 0|\, D\, {\cal H}_w {\cal H}_w \,D\, |0 \rangle\,.
\eeq
Since the operator $D$ is of the form $\bar cu$, it transforms in the
fundamental representation of $SU(3)_F$, which we will represent with a
lower index, $D_i$.  We use a convention in which the correspondence between 
matrix indexes and quark flavors is $(1,2,3)=(u,d,s)$.  The only nonzero 
element of $D_i$ is $D_1=1$.  The $\Delta C=-1$
part of the weak Hamiltonian has the flavor structure $(\bar q_ic)(\bar
q_jq_k)$, so its matrix representation is written with a fundamental
index and two antifundamentals, $H^{ij}_k$.  This operator is a sum of irreps
contained in the product $3 \times \3bar \times \3bar =
\15bar + 6 + \3bar + \3bar$.  In the limit in which the third generation is
neglected, $H^{ij}_k$ is traceless, so only the $\15bar$ 
and 6 representations appear.  That is, the $\Delta C=-1$ part of ${\cal H}_w$ 
may be decomposed as ${1\over2} (\cO_{\15bar} + \cO_6)$.

Since we are interested in $SU(3)_F$ breaking, let is introduce it through the quark 
mass operator ${\cal M}$, whose matrix representation is $M^i_j={\rm diag}(m_u,m_d,m_s)$.  
We set $m_u=m_d=0$ and let $m_s\ne0$ be the only $SU(3)$ violating parameter.  
All nonzero matrix elements built out of $D_i$, $H^{ij}_k$ and $M^i_j$ must be 
$SU(3)_F$ singlets.

We now prove that $\DDbar$ mixing arises only at second order in
$SU(3)$ violation, by which we mean second order in $m_s$.  First, we note that
the pair of $D$ operators is symmetric, and so the product $D_iD_j$
transforms as a 6 under $SU(3)$.  Second, the pair of ${\cal H}_w$'s is also symmetric, 
and the product $H^{ij}_kH^{lm}_n$ is in one of the reps which
appears in the product
\beqa
\left[ (\15bar+6)\times(\15bar+6) \right]_S =
    (\15bar\times\15bar)_S +(\15bar\times 6)+(6\times 6)_S ~~~~~~~~~\\*
= (\60bar+\24bar+15+15'+\sixbar) + (42+24+15+\sixbar+3)
    + (15'+\sixbar)\,. \nonumber
\eeqa
A direct computation shows that only three of these
representations actually appear in the decomposition of ${\cal H}_w{\cal H}_w$.  They 
are the $\60bar$, the 42, and the $15'$ 
\beqa
    DD = {\cal D}_6\,, ~~~~
    {\cal H}_w {\cal H}_w = \cO_{\60bar}+\cO_{42}+\cO_{15'}\,,
\eeqa
where subscripts denote the representation of $SU(3)_F$.
Since there is no $\sixbar$ in the decomposition of ${\cal H}_w{\cal 
H}_w$, there is no $SU(3)$ singlet which can be made with ${\cal D}_6$,  
and no $SU(3)$ invariant matrix element of the form (\ref{melm}) can be formed.  
This is the well known result that $D^0-\D0bar$ mixing is 
{\it prohibited by $SU(3)$ symmetry}.
Now consider a single insertion of the $SU(3)$ violating spurion ${\cal M}$.
The combination ${\cal D}_6{\cal M}$ transforms as $6\times
8=24+\15bar+6+\3bar$. There is still no invariant to be made
with ${\cal H}_w{\cal H}_w$, thus $D^0-\D0bar$ mixing is {\it not 
induced at first order in $SU(3)_F$ breaking}.
With two insertions of ${\cal M}$, it becomes possible to make an $SU(3)_F$
invariant.  The decomposition of ${\cal D}{\cal M}{\cal M}$ is
\beqa
6\times(8\times 8)_S &=& 6\times(27+8+1) \\
    &=& (60+\42bar+24+\15bar+\15bar'+6) + (24+\15bar+6+\3bar) + 6\,.
\nonumber
\eeqa
There are three elements of the $6\times 27$ part which can give invariants
with  ${\cal H}_w{\cal H}_w$.  Each invariant yields a contribution to 
$D^0-\D0bar$ mixing proportional to $s_1^2m_s^2$. Thus, $\DDbar$ mixing arises 
only at {\it second order} in the $SU(3)$ violating parameter $m_s$~\cite{Falk:2001hx}, 
in the Standard Model $x$ and $y$ are generated only at second order in $SU(3)_F$ 
breaking, 
\beq
x\,,\, y \sim \sin^2\theta_C \times [SU(3) \mbox{ breaking}]^2\,,
\eeq
where $\theta_C$ is the Cabibbo angle.  This result should be reproduced in all 
explicit calculations of $\DDbar$ mixing parameters.

The use of the OPE relies on local quark-hadron duality, and on expansion parameter $\Lambda/m_c$ being 
small enough to allow truncation of the series after the first few terms. Let us see what one can expect at 
leading order in $1/m_c$ expansion, i.e. assuming that the integrals in Eq.~(\ref{XandY}) are dominated by 
the short distances. The leading-order result is then generated by calculating the usual box diagram with 
intermediate $s$ and $d$ quarks.
 
Unitarity of the CKM matrix assures that the leading-order, mass-independent contribution due to $s$-quark is
completely cancelled by the corresponding contribution due to a $d$-quark. A non-zero contribution 
can be obtained if the mass insertions are added on each quark line in the box diagram. However, adding only
one mass insertion flips the chirality of the propagating quarks, from being left-handed to right-handed. 
This does not give a contribution to the resulting amplitude, as right-handed quarks do not
participate in weak interaction. Thus, a second mass insertion is needed on each quark line. Neglecting $m_d$
compared to $m_s$ we see that the resulting contribution to $x_D$ is 
${\cal O}(m_s^2\times m_s^2) \sim {\cal O}(m_s^4)$! It is easy to convince yourself that $y_D$ has additional 
$m_s^2$ suppression due to on-shell propagation of left-handed quarks emitted from a spin-zero meson,
which brings total suppression of $y_D$ to ${\cal O}(m_s^6)$!  An explicit calculation of the leading order 
mixing amplitude, as well as perturbative QCD corrections to it~\cite{Golowich:2005pt} agrees precisely 
with the hand-waving arguments above. Clearly, leading order contribution in $1/m_c$ gives ``too much" 
of $SU(3)_F$ suppression compared to the theorem that was proven above~\cite{Falk:2001hx}.

Somewhat surprisingly, the resolution of this paradox follows from considerations of higher-order corrections in 
$1/m_c$~\cite{Inclusive}. Among many higher-dimensional operators that encode $1/m_c$ corrections to the
leading four-fermion operator contribution, there exists a class of operators that result from chirality-flipping 
interactions with background quark condensates. These interactions do not bring additional powers of 
light quark mass, but are suppressed by powers of $\Lambda_{QCD}/m_c$, which is not a very small number.
The leading ${\cal O}(m_s^2)$ order of $SU(3)_F$ breaking is obtained from matrix elements of dimension  twelve
operators that are suppressed by $(\Lambda_{QCD}/m_c)^6$ compared to the parametrically-leading 
contribution in $1/m_c$ expansion~\cite{Inclusive}! As usual in OPE calculation, proliferation of the number of 
operators at higher orders (over 20) makes it difficult to pinpoint the precise value of the effect.

\section{Threshold effects in OPE and exclusive approaches to 
calculation of mixing parameters}

There are several concerns that one need to deal with when calculating $\DDbar$ mixing using OPE-based methods.
First, the numerically leading order effect comes from dimension twelve operators. Quark-hadron duality was never 
checked for such case~\cite{Shifman:2000jv}. Second, the number of matrix elements of operators is very large. It is 
not clear how to properly combine uncertainties associated with computations of those matrix elements. Third concern, 
which is also related to the issue of quark-hadron duality, regards the proper way of dealing with hadronic thresholds in 
OPE framework. 

Let us concentrate on calculation of $y_D$. In order to illustrate the issue of hadronic thresholds, one needs to 
recall that heavy quark operator expansion is really an expansion in the energy released in the process of the decay.
In D-decays this energy is not always large: for example, for $KKK$ intermediate state the energy released in the decay
is $E_r \sim m_D - 3 m_K \sim {\cal O}(\Lambda_{QCD})$, which is by no means large. In the limit $m_c \to \infty$ one
immediately sees that $m_c \gg M_{\mbox{\tiny intermediate state}}$, so $E_r \sim m_c$. This is why this approach 
works very well for B-decays, but might have issues for charm. It is clear that more fork is needed to understand 
range of applicability of OPE methods to $\DDbar$ mixing~\cite{Lenz:2013aua}.  

It is possible to calculate $\DDbar$ mixing rates by dealing explicitly with hadronic intermediate states 
which result from every common decay product of $D^0$ and $\D0bar$~\cite{Exclusive}. 
In the $SU(3)_F$ limit, these contributions cancel when one sums over complete $SU(3)$ multiplets in 
the final state.  The cancellations depend on $SU(3)_F$ symmetry both in the decay matrix 
elements and in the final state phase space. While there are $SU(3)$ violating
corrections to both of these, it is difficult to compute the $SU(3)_F$ violation in the matrix 
elements in a model independent manner. As experimental data on nonleptonic decay
rates becomes better and better, it is possible to use it to calculate $y_D$ by
directly inputing it into Eq.~(\ref{XandY}),
\beq
y_D = {1\over\Gamma} \sum_n \int [{\rm P.S.}]_n\,
    \langle \D0bar|\,{\cal H}_w\,|n \rangle \langle n|\,{\cal H}_w\,|D^0
\rangle\,,
\eeq
where the sum is over distinct final states $n$ and the integral is over
the phase space for state $n$.  Alternatively, with 
some mild assumptions about the momentum dependence of the matrix elements, the 
$SU(3)_F$ violation in the phase space depends only on the final particle masses and 
can be computed~\cite{Falk:2001hx}. It was shown that this source of $SU(3)_F$ 
violation can generate $y_D$ and $x_D$ of the order of a few percent. The calculation of 
$x_D$ relies on further model-dependent assumptions about off-shell behavior 
of decay form-factors~\cite{Falk:2001hx}. Restricting the sum over all final states to 
final states $F$ which transform within a single $SU(3)_F$ multiplet $R$, the result is
\beq
 y_D =   {1\over2\Gamma}\, \langle\D0bar|\,{\cal H}_w
     \bigg\{ \eta_{CP}(F_R)\sum_{n\in  F_R}
    |n\rangle \rho_n\langle n| \bigg\} {\cal H}_w\,|D^0\rangle\,,
\eeq
where $\rho_n$ is the phase space available to the state $n$, 
$\eta_{CP}=\pm1$~\cite{Falk:2001hx}.  In the
$SU(3)_F$ limit, all the $\rho_n$ are the same for $n\in F_R$, and the quantity 
in braces above is an $SU(3)_F$ singlet.  Since the $\rho_n$ depend only on the 
known masses of the particles in the state $n$, incorporating the true values
of $\rho_n$ in the sum is a calculable source of $SU(3)_F$ breaking.

This method does not lead directly to a calculable contribution to $y$,
because
the matrix elements $\langle n|{\cal H}_w|D^0\rangle$ and
$\langle\D0bar|{\cal H}_w|n\rangle$
are not known.  However, $CP$ symmetry, which in the Standard Model and
almost all scenarios of new physics is to an excellent approximation
conserved in $D$ decays, relates $\langle\D0bar|{\cal H}_w|n\rangle$ to
$\langle
D^0|{\cal H}_w|\overline{n}\rangle$. Since $|n\rangle$ and
$|\overline{n}\rangle$ are
in a common $SU(3)_F$ multiplet,  they are determined by a single effective
Hamiltonian. Hence the ratio
\beqa\label{yfr}
   y_{F,R} &=& {\sum_{n\in F_R} \langle\D0bar|\,{\cal H}_w|n\rangle \rho_n
    \langle n|{\cal H}_w\,|D^0\rangle \over
    \sum_{n\in F_R} \langle D^0|\,{\cal H}_w |n\rangle \rho_n
    \langle n|{\cal H}_w\,|D^0\rangle} \nonumber \\
   &=& {\sum_{n\in F_R} \langle\D0bar|\,{\cal H}_w|n\rangle \rho_n
    \langle n|{\cal H}_w\,|D^0\rangle \over \sum_{n\in F_R}
   \Gamma(D^0\to n)} 
\eeqa
is calculable, and represents the value which $y_D$ would take if elements
of $F_R$ were the only channel open for $D^0$ decay.  To get a true
contribution
to $y_D$, one must scale $y_{F,R}$ to the total branching  ratio to all the
states in $F_R$.  This is not trivial, since a given physical final state
typically decomposes into a sum over more than one multiplet $F_R$.  The
numerator of $y_{F,R}$ is of order $s_1^2$ while the  denominator is of
order 1, so with large $SU(3)_F$ breaking in the phase space the natural size 
of $y_{F,R}$ is 5\%.
Indeed, there are other $SU(3)_F$ violating effects, such as in matrix elements 
and final state interaction phases.  Here we assume that
there is no cancellation with other sources of $SU(3)_F$ breaking, or
between the various multiplets which occur in $D$ decay, that would
reduce our result for $y$ by an order of magnitude.
This is equivalent to assuming that the $D$ meson is not
heavy enough for duality to enforce such cancellations. Performing the 
computations of $y_{F,R}$, we see~\cite{Falk:2001hx} that effects at the level 
of a few percent are quite generic.  
Our results are  summarized in Table~\ref{ytwobody}. Then, $y_D$ can be
formally constructed from the individual $y_{F,R}$ by
weighting them by their $D^0$ branching ratios,
\beq\label{ycombine}
    y_D = {1\over\Gamma} \sum_{F,R}\, y_{F,R}
    \bigg[\sum_{n\in F_R}\Gamma(D^0\to n)\bigg]\,.
\eeq
However, the data on $D$ decays are neither abundant nor precise enough
to disentangle the decays to the various $SU(3)_F$ multiplets, especially
for the three- and four-body final states.  Nor have we computed  $y_{F,R}$ for
all or even most of the available representations.   Instead, we can only
estimate individual contributions to $y$ by  assuming that the representations
for which we know $y_{F,R}$ to be  typical for final states with a given
multiplicity, and then to scale to  the total branching ratio to those final states.
The total branching  ratios of $D^0$ to two-, three- and four-body final
states can be extracted from the Review of Particle Physics. Rounding to the 
nearest 5\% to emphasize 
the uncertainties in these numbers, we conclude that the branching fractions for 
$PP$, $(VV)_{\mbox{$s$-wave}}$, $(VV)_{\mbox{$d$-wave}}$ and $3P$ approximately 
amount to 5\%, while the branching ratios for $PV$ and $4P$ are of the order 
of 10\%~\cite{Falk:2001hx}.

\begin{table}
\begin{center}
\begin{tabular}{|@{~~~}lc|c|c|} \hline
\multicolumn{2}{|c|}{~~Final state representation~~~}  &
     ~~~$y_{F,R}/s_1^2$~~~ & ~~~$y_{F,R}\ (\%)$~~~  \\ \hline\hline
    $PP$  &  $8$  &  $-0.0038$ & $-0.018$  \\
    &  $27$  &  $-0.00071$  & $-0.0034$ \\ \hline
    $PV$  &  $8_A$  &  $0.032$ & $0.15$\\
    &  $8_S$  &  $0.031$  & $0.15$ \\
    &  $10$  &  $0.020$ & $0.10$ \\
    &  $\overline{10}$  &  $0.016$ & $0.08$ \\
    &  $27$  &  $0.04$  & $0.19$\\ \hline
    $(VV)_{\mbox{$s$-wave}}$  &  $8$  &  $-0.081$ & $-0.39$ \\
    &  $27$  &  $-0.061$ & $-0.30$\\
    $(VV)_{\mbox{$p$-wave}}$  &  $8$  &  $-0.10$ & $ -0.48$\\
    &  $27$  &  $-0.14$ & $-0.70$ \\
    $(VV)_{\mbox{$d$-wave}}$  &  $8$  &  $0.51$ & $2.5$ \\
    &  $27$  &  $0.57$  & $2.8$\\ \hline
$(3P)_{\mbox{$s$-wave}}$        &  $8$  &  $-0.48$  & $-2.3$\\
    &  $27$  &  $-0.11$  & $-0.54$ \\
$(3P)_{\mbox{$p$-wave}}$        &  $8$  &  $-1.13$  & $-5.5$ \\
    &  $27$  &  $-0.07$   & $-0.36$  \\
$(3P)_{\mbox{form-factor}}$     &  $8$  &  $-0.44$  & $-2.1$\\
    &  $27$  &  $-0.13$ & $-0.64$ \\ \hline
$4P$  &  $8$  &  $3.3$ & $16$  \\
    &  $27$  &  $2.2$  & $11$ \\
    &  $27'$  &  $1.9$ & $9.2$ \\ \hline
\end{tabular} \vspace{4pt}
\caption{Values of $y_{F,R}$ for some two-, three-, and four-body final states.}
\label{ytwobody}
\end{center}
\end{table}

It can be easily seen that there are terms in Eq.~(\ref{ycombine}), like nonresonant $4P$, 
which could make contributions  to $y_D$ at the level of a percent or larger.  
There, the rest masses of the final state particles take up most of the available 
energy, so phase space differences are very important. One can see
that $y_D$ on the order  of a few percent is completely natural, and that anything 
an order of magnitude smaller would require significant  cancellations which do not
appear naturally in this framework.  The normalized mass difference, $x_D$, can
then be calculated via a dispersion relation
\beq
x_D = - \frac{1}{\pi}\ \mbox{P} \int_{2 m_\pi}^\infty dE \frac{y_D(E)}{E-m_D}
\eeq
that additionally contain guesses on the off-shell behavior of hadronic form-factors 
in $y_D(E)$~\cite{Falk:2001hx}. Here P denotes principal value.
The result of the calculation yields $x_D \sim {\cal O}(1\%)$~\cite{Falk:2001hx}.

Since experimental data on nonleptonic decays of charmed mesons improved significantly in 
the past several years, it can be used to estimate some contributions~\cite{Cheng:2010rv}. 
For example, concentrating on the $\pi\pi$, $KK$, and $\pi K$ intermediate states,  
\bea\label{y2}
    y_{2D} &=& \mbox{Br} (D^0\to K^+K^-) + \mbox{Br} (D^0\to \pi^+\pi^-)  
    \nonumber \\
    &-& 2\cos\delta \sqrt{\mbox{Br} (D^0\to K^+\pi^-) \mbox{Br} (D^0\to \pi^+K^-)}
\eea
The PDG values~\\cite{Beringer:1900zz} for the branching ratios above are known quite well 
for the purpose of calculation of $y_{2D}$,
\bea\label{BraRat}
    \mbox{Br} (D^0\to K^+K^-) &=& (3.96\pm0.00) \times 10^{-3}, 
    \nonumber \\
    \mbox{Br} \ (D^0\to \pi^+\pi^-)  &=& (1.401\pm0.027) \times 10^{-3},
    \nonumber \\
    \mbox{Br} (D^0\to K^+\pi^-) &=& (3.88\pm0.05) \times 10^{-2},  
    \\
    \mbox{Br} (D^0\to \pi^+K^-) &=& (1.31\pm0.08) \times 10^{-4}.
    \nonumber
\eea
Notice that $\cos\delta$ is not known well. Its value however is very important for numerical
calculation of Eq.~(\ref{y2}), as large cancellations (between the first and the second lines
of that equation) are expected. Taking the $U$-spin limit $\cos\delta=1$~\cite{Falk:1999ts},
one arrives at the contribution $y_{2D} = (0.85 \pm 0.17)\times 10^{-3}$. Unfortunately, 
other branching ratios, especially for three or four body decays, are not known that well.
Therefore, saturating Eq.~(\ref{yfr}) with experimental data would only be sensitive to the
values of experimental uncertainties of measurements of branching ratios, not to the true
size of the effect. 

It must be pointed out that similar calculations of $y_D$ have been recently carried out 
using the simpler language of $U$-spin with consistent results~\cite{Gronau:2012kq}.

\section{Outlook}

The calculation of $\DDbar$ mixing is a challenging theoretical exercise. 
It is not clear if brute-force improvements of the calculations would result in 
much more precise results. 

However, a glimpse of hope for yet another approach have recently been identified. 
Probably not surprisingly, it came from lattice QCD calculations, which usually
shined away from the calculations of non-leptonic decay amplitudes. 
It remains to be seen if multichannel generalizations of Lellouch-Luscher 
approaches~\cite{Lellouch:2000pv} to calculations of weak matrix 
elements will be successful in calculating non-leptonic decay rates of charm 
mesons~\cite{Hansen:2012tf}. Yet, this approach will certainly 
have impact on charm physics and, in particular, on calculations of $\DDbar$ mixing rate.

\Acknowledgements
It is my pleasure to thank S. Bergmann, A. Falk, E. Golowich, Y. Grossman, J. Hewett, 
Z. Ligeti, Y. Nir, and S. Pakvasa for collaborations on the papers related to this review talk. 
I would like to thank the organizers for the invitation to the wonderfully organized workshop.

\end{document}